# Translating the Force Concept Inventory in the age of AI

Marina Babayeva
Department of Physics Education, Charles University,
Prague, Czech Republic
marina.babayeva@matfyz.cuni.cz
ORCID 0000-0002-5778-9722

Justin Dunlap
Portland State University, Portland, USA

Marie Snětinová
Department of Physics Education, Charles University,
Prague, Czech Republic

Ralf Widenhorn
Portland State University, Portland, USA

**Abstract**

We present a study that translates the Force Concept Inventory (FCI) using OpenAI's GPT-4o and assess the specific difficulties of translating a scientific-focused topic using Large Language Models (LLMs). The FCI is a physics exam meant to evaluate outcomes of a student cohort before and after instruction in Newtonian physics. We examine the problem-solving ability of the LLM in both the translated document and the translation back into English, detailing the language-dependent issues that complicate the translation. While ChatGPT performs remarkably well on answering the questions in both the translated language as well as the back-translation into English, problems arise with language-specific nuances and formatting. Pitfalls include words or phrases that lack one-to-one matching terms in another language, especially discipline-specific scientific terms, or outright mistranslations. Depending on the context, these translations can result in a critical change in the physical meaning of the problem. Additionally, issues with question numbering and lettering are found in some languages. The issues around the translations of numbering and lettering provide insight into the abilities of the LLM and suggest that it is not simply relying upon FCI questions that may have been part of the LLM's training data to provide answers. These findings underscore that while LLMs can accelerate multilingual access to educational tools, careful review is still needed to ensure fidelity and clarity in translated assessments. LLMs provide a new opportunity to expand educational tools and assessments. At the same time, there are unique challenges using LLMs to facilitate translations that this case study examines in detail.

1. Introduction

**Physics concept inventories**

Artificial intelligence (AI) and in particular, Large Language Models (LLMs) have created new opportunities and challenges for physics educators since the widespread use of these tools has become available (Küchemann, Avila, et al., 2024; Yusuf et al., 2024, Rahman & Watanobe, 2023). Physics education researchers, like most scientific communities, share much of their results in English. Therefore, standard concept assessments that have been research-validated in different environments often exist only in English. Some assessments have been translated into other languages by physics experts with appropriate language skills. However, the majority of physics concept inventories do not exist in a large set of languages. The Force Concept Inventory (FCI) is a widely researched and commonly used tool to measure the outcomes of instructing students in Newtonian physics (Carleschi et al., 2022; Stoen et al., 2020; Koca & Suleiman, 2019; Laverty & Caballero, 2018) and LLM's ability to solve the FCI has been studied previously (West, 2023). There are 33 translations of the FCI shared on PhysPort (*PhysPort Assessments: Force Concept Inventory*, 2025), by far the most of any physics concept inventories, but there are many languages that are not covered.

Here, we will explore some of the opportunities and challenges for this well-studied assessment, from the extraction of the digital file to the machine translation and assessment of machine translation. For all translations, we will try to use the LLM to provide us with information on how well the translation kept the physics context by asking the LLM to answer the concept inventory before and after translation. This has inherent limitations, as currently, only a human physics content expert with appropriate languages in both English and the language the assessment is translated into can and should fully validate a translation. However, we argue that using AI can give some indication of the validity of a translation. We will use our knowledge of some languages (German, Russian, and Czech) and have consulted with physicists in other languages (Marathi and Hindi) to discuss what aspects can impact the quality of translations of this physics concepts inventory and the implications of translating other physics educational content.

**AI in physics education**

Artificial Intelligence (AI) is increasingly shaping the educational landscape, offering tools that support teaching, assessment, and content adaptation. LLMs have demonstrated their capabilities in various educational applications, including automated feedback, teaching assistance, and resource creation (Kregear et al., 2025; Kortemeyer, 2024; Steinert et al., 2024; Küchemann et al., 2023). Several studies have explored the use of AI in education, highlighting its potential to assist teachers as interlocutors, with content creation and evaluation, as well as providing feedback for written assignments (e.g., Banihashem et al., 2024; Jeon & Lee, 2023; Shahriar &

Hayawi, 2023). AI-based tools are being explored for grading, solving problems, and acting as virtual tutors (Kortemeyer et al., 2024; Yeadon & Hardy, 2024; Steinert, Avila, et al., 2024). However, challenges remain in ensuring that AI-generated responses maintain scientific accuracy and pedagogical effectiveness. Prompt engineering and understanding the inner workings of LLMs is important to guide its effective and responsible use in physics education (Polverini & Gregorcic, 2024b). Recent advancements in LLMs, such as OpenAI's GPT-o3 and GPT-5, Google's AlphaGeometry, and DeepSeek's R1, have shown that even higher-order thinking tasks are no longer insurmountable challenges for AI models. These systems exhibit remarkable proficiency in complex problem-solving, suggesting a growing potential for AI to support advanced conceptual understanding in physics education.

**Languages and physics**

While LLMs have been evaluated in their ability to generate and assess physics-related content, their performance can vary across languages due to disparities in training data and linguistic complexity (Kortemeyer et al., 2025, GPT-4, 2023; Nicholas & Bhatia, 2023). Research shows that LLMs trained predominantly on English materials may introduce cultural biases when generating responses in other languages (Pava et al., 2025; Tao et al., 2024; Kwak & Pardos, 2024; Karinshak et al., 2024). This phenomenon has been observed in studies analyzing AI-generated feedback in multilingual educational settings, highlighting concerns about consistency and accuracy (Cohere For AI, 2024). Furthermore, differences in syntax, vocabulary, and grammar affect how AI interprets and translates educational content across languages (Feng et al., 2024). These linguistic disparities pose challenges for AI applications in physics education, where precise terminology and conceptual clarity are critical.

LLM and machine translation (MT) technologies may be a solution to overcome some of the challenges of sharing educational resources internationally. Studies show that MT tools are widely used by students for quick translations of academic texts, especially in non-language-focused disciplines, offering speed and accessibility but often lacking contextual nuance and accuracy (Powell et al., 2022; Steigerwald et al., 2022). In contrast, LLMs demonstrate superior performance in translating complex academic content, capturing tone, semantics, and cultural subtleties more effectively. LLMs are used to generate high-quality bilingual corpora and enhance translation competence, though challenges remain around overreliance, ethical use, and the need for teacher training (Aleedy et al., 2025; Nguyen et al., 2025; Mohsen, 2024). Overall, the shift from MT to LLM translation marks a move from basic utility to intelligent, context-aware educational support. Researchers present LLM debiasing techniques for improving the performance across different languages, including low-resource languages (e.g., Singh et al., 2024; Dang et al., 2024; Reusens et al., 2023). Neural network machine translation models can capture nuanced meanings across languages, preserving the educational value of materials while making them accessible to different linguistic audiences (e.g., Costa-jussà et al., 2024; Weglot & Nimdzi Insights, 2022; Muftah, 2022; Popel et al., 2020). These systems increasingly recognize idiomatic expressions and can adapt them appropriately for target languages. Therefore, our research focuses on LLM-based translation.

Despite these promising capabilities, significant challenges remain in implementing AI solutions for educational resource sharing. The LLMs still struggle with cross-lingual knowledge transfer (Chua et al., 2024). Therefore, quality assurance processes must be developed to ensure that AI-transformed educational materials maintain accuracy and effectiveness across cultures and languages. The problem was acknowledged by the European Union with the project Open Euro LLM (*Open Euro LLM*, n.d.). Human oversight remains essential, as fully automated solutions may miss subtle cultural nuances, contextual information, and domain-specific vocabulary (e.g., Kocmi et al., 2024; Naveen & Trojovský, 2024), which in pedagogical considerations may affect learning outcomes. AI and machine translation technologies offer powerful tools for breaking down barriers to international educational resource sharing, but they require thoughtful implementation with ongoing human guidance to achieve their full potential.

**Research Questions**

This case study aims to investigate the performance and challenges of using an LLM for translating and assessing a multiple-choice physics concept inventory across multiple languages. Specifically, we seek to answer the following research questions:

1. What are the challenges to machine transcribing and translating an English version physics multiple-choice concept inventory from a PDF into a form that can be easily processed by an LLM and used by human instructors?

2. Are there language-based differences from the translations that impact the LLM's problem-solving ability of the concept inventory?

To address the first research question, we extracted text from PDF documents through LLM transcription, followed by manual corrections and modifications of question sequences. The outcome of this step was a complete textual representation of the FCI in English, along with original image files. The second step entailed generating LLM-based translations of the FCI into 53 selected languages and translations back to English. We will discuss the challenges associated with the formatting of the digital file, special characters like subscripts, equations, the placement of figures, and sequences of questions that are connected. For selected questions and languages, we will discuss the quality of the translations. Additionally, we provide an example of comparison between LLM translation versus physics expert and native speaker translation.

To address the second research question, assessing language-specific differences, the translated and back-translated questions were submitted to the LLM for solving. This step required language-specific prompting and postprocessing, particularly for handling non-Latin scripts. The final output comprised the model-generated multiple-choice selections and explanations in the respective languages. We will discuss if and what the LLM responses can tell us about the translations. For selected languages, we will discuss the LLM answers directly by analyzing the responses.

## 2. Method

**Transcription and Translation**

The first step in data preparation involved transcribing the FCI from its English version available as a PDF on PhysPort. Transcription was automated using OpenAI's GPT-4o-mini model (this is the only place in this study where the mini version was used, it was done for cost reasons) with a request to individually transcribe each question in the PDF without adding any additional text While generally successful, the model occasionally omitted parts of questions, misnumbered items, or introduced formatting inconsistencies (e.g., missing subscripts). All transcriptions were manually reviewed and corrected to correspond to the original document.

Following transcription, questions were edited to ensure that each question can be solved independently. We modified items that relied on shared figures or an introductory text for multiple questions. For instance, Questions 8–11 originally referenced a common setup (introductory text and image); we included this setup directly into each question to make each self-contained. Where necessary, multiple figures were combined into one image to reflect the full visual context. Final question texts and graphics were stored in UTF-8 encoded comma-separated files (CSV) for the text data and Portable Network Graphics (PNG) images for the figures.

The full set of FCI questions was then translated into 53 languages using OpenAI's gpt-4o-2024-11-20 model (unless indicated otherwise, this version was for all tasks used in this study). Languages were selected based on availability in the *langdetect* Python library (*Langdetect*, 2021), and include a diverse mix of languages from around the globe (see Table 3 for list of languages). Langdetect was used to confirm that a translation was in the target language but did not provide insight into the accuracy of the translation. During translation, the role was set in English to *"You are a helpful assistant"* using the default temperature of 0.7. All translation prompts specified that responses should contain only the translated text, avoiding any meta-comments. A total of 53 languages with 30 questions per test resulted in 1,590 questions. Manual review identified a few issues: two questions in two languages had incomplete translations (missing either the question or the answer choices), and one translation in Malayalam included a nonsensical combination of Malayalam and English that was identifiable even to non-speakers of Malayalam. These were corrected either manually or by reprocessing with the LLM. Additionally, we observed that in 17 non-Latin languages, localized characters were used in answer choices; these questions were kept unchanged. All of the outputs from the translations were exported to a CSV file.

Each language version was also back-translated into English. Back-translations revealed further problems, including missing content in 30 questions in 20 languages and two cases of added meta-text, such as "*The phrase translates to:*" and "*Official translation:*". After reprocessing, 9 questions still showed missing content; these were manually edited. While the final translated versions were used for problem-solving, the English back-translations served to validate translation consistency and supported vocabulary analysis across languages.

Additionally, all prompts given to the LLM to solve the problems were translated into the corresponding language prior to having the LLM solve the questions in order to have the exchange with the LLM in only one language during the problem-solving request.

**Problem-solving**

After translation, each version of the FCI was submitted to the GPT-4o model (gpt-4o-2024-11-20) for problem-solving using corresponding prompts in the target language, translated by the LLM. The model role was set in English to a neutral statement, *"You are an AI that can analyze and describe images,"* and relevant question figures were attached where applicable. We used two separate prompts per question: one requesting a full explanation based on the provided figure, and another asking for the final answer as a single letter. In English, they were:

*"Solve the following physics problem. If there are one or more figures, describe their content and discuss how you used them to solve the problem."*

and

*"Restate the correct multiple-choice answer. Make sure that your answer agrees with your response and reasoning in the previous question. Your answer should only be one letter (e.g. "B") and no additional text."*

Translated prompts were verified via back-translation to ensure they remained semantically aligned with the original English instructions. For quality assurance, we verified prompt clarity in languages familiar to our team (English, German, Russian, and Czech) and manually checked all back-translations.

Using *langdetect*, we were able to confirm that all LLM responses were in the target language. This is different from the Kortemeyer study, where the problems were given to the LLM in different languages, but the prompts were only given in English. In that study, it was found that the responses from the LLM included a mix of English and the language of the provided problems (Kortemeyer et al., 2025). Responses were then post-processed to standardize multiple-choice answers and address recurring issues, such as the inclusion of extra explanatory text or non-Latin characters in answer labels. Additionally, as a baseline, the full English version of the FCI was solved five times each using two models: gpt-4o-2024-11-20 and o1-2024-12-17 as the most advanced models available in the general purpose and reasoning areas at the time of the research. These results served as reference points to interpret multilingual outputs. By comparing reasoning and answer consistency across languages, we indirectly assessed how well the translated versions preserved the solvability and intended meaning of the original questions.

3. Results

In this section, we present the results of problem-solving performance on translated versions of the FCI using LLMs. First, we will take a closer look at how multiple-choice options were represented across different languages. In English-language multiple-choice concept inventories, such as the FCI, Latin letters are typically used to label response options. During the translation and problem-solving process, we observed that most language versions retained Latin letters for these options. However, in several translations using non-Latin scripts, local characters were also used as multiple-choice selectors. The languages for which the LLM used non-Latin letters in some or all of their multiple-choice selectors are shown in Table 1. As we will discuss, this mix of Latin and non-Latin multiple-choice options can lead to issues.

| Language | Non-Latin MC in the question translation | Non-Latin MC in the answer |
|---|---|---|
| Arabic | 20 | 30 |
| Bengali | 11 | 10 |
| Bulgarian | 2 | 30 |
| Greek | 22 | 26 |
| Hebrew | 14 | 30 |
| Hindi | 5 | 4 |
| Kannada | 6 | 1 |
| Macedonian | 16 | 30 |
| Marathi | 4 | 14 |
| Nepali | 15 | 10 |
| Persian | 4 | 20 |
| Punjabi | 4 | - |
| Telugu | 1 | - |
| Thai | 9 | 2 |
| Urdu | 4 | 4 |

*Table 1* - *Languages for which a non-Latin script was used for some of the multiple-choice (MC) selectors either during the translation of the question or the LLM's response. The number of questions out of 30 with non-Latin letters is shown.*

Table 2 shows the distribution of the answers from LLM among correct answers during the problem-solving stage, the most frequent wrong answer, and second most common wrong answer. It also shows three different question categories: text-only, image-optional, and image-required questions. Among the 30 FCI questions, we categorized 10 questions as text-only, 9 questions contained images but were non-essential for solving the problem and were categorized as image-optional, and 11 questions relied on crucial visual components for problem-solving and were categorized as image-required. For the questions the model struggled with, typically those requiring interpretation of images, its incorrect answers were not evenly spread across the multiple-choice options, but instead clustered around specific distractors. The model clearly gravitated toward particular incorrect options.

| # | Type | Languages | | | Back translations | | | English GPT-4o | | | English GPT-o1 | | |
|---|---|---|---|---|---|---|---|---|---|---|---|---|---|
| | | Correct option | Most frequent wrong option | 2nd most frequent wrong option | Correct option | Most frequent wrong option | 2nd most frequent wrong option | Correct option | Most frequent wrong option | 2nd most frequent wrong option | Correct option | Most frequent wrong option | 2nd most frequent wrong option |
| 1 | 0 | **100%** | 0% | 0% | **100%** | 0% | 0% | **100%** | 0% | 0% | **100%** | 0% | 0% |
| 2 | 0 | **100%** | 0% | 0% | **100%** | 0% | 0% | **100%** | 0% | 0% | **100%** | 0% | 0% |
| 3 | 0 | **100%** | 0% | 0% | **100%** | 0% | 0% | **100%** | 0% | 0% | **100%** | 0% | 0% |
| 4 | 0 | **100%** | 0% | 0% | **100%** | 0% | 0% | **100%** | 0% | 0% | **100%** | 0% | 0% |
| 5 | 2 | **100%** | 0% | 0% | **100%** | 0% | 0% | **100%** | 0% | 0% | **100%** | 0% | 0% |
| 6 | 2 | **96%** | 4% | 0% | **96%** | 2% | 2% | **100%** | 0% | 0% | **100%** | 0% | 0% |
| 7 | 2 | **4%** | 96% | 0% | **0%** | 98% | 2% | **0%** | 100% | 0% | **0%** | 80% | 20% |
| 8 | 2 | **38%** | 49% | 9% | **34%** | 57% | 8% | **60%** | 60% | 40% | **80%** | 20% | 0% |
| 9 | 1 | **92%** | 8% | 0% | **85%** | 15% | 0% | **100%** | 0% | 0% | **100%** | 0% | 0% |
| 10 | 1 | **100%** | 0% | 0% | **100%** | 0% | 0% | **100%** | 0% | 0% | **100%** | 0% | 0% |
| 11 | 1 | **98%** | 2% | 0% | **98%** | 2% | 0% | **100%** | 0% | 0% | **100%** | 0% | 0% |
| 12 | 2 | **6%** | 89% | 6% | **2%** | 92% | 4% | **0%** | 80% | 20% | **20%** | 80% | 0% |
| 13 | 0 | **100%** | 0% | 0% | **100%** | 0% | 0% | **100%** | 0% | 0% | **100%** | 0% | 0% |
| 14 | 2 | **25%** | 75% | 0% | **32%** | 62% | 6% | **0%** | 100% | 0% | **0%** | 100% | 0% |
| 15 | 1 | **100%** | 0% | 0% | **100%** | 0% | 0% | **100%** | 0% | 0% | **100%** | 0% | 0% |
| 16 | 1 | **100%** | 0% | 0% | **100%** | 0% | 0% | **100%** | 0% | 0% | **100%** | 0% | 0% |
| 17 | 1 | **100%** | 0% | 0% | **100%** | 0% | 0% | **100%** | 0% | 0% | **100%** | 0% | 0% |
| 18 | 2 | **96%** | 4% | 0% | **100%** | 0% | 0% | **100%** | 0% | 0% | **100%** | 0% | 0% |
| 19 | 2 | **11%** | 57% | 17% | **9%** | 62% | 15% | **0%** | 60% | 40% | **100%** | 0% | 0% |
| 20 | 2 | **2%** | 83% | 11% | **4%** | 77% | 17% | **0%** | 60% | 40% | **20%** | 60% | 20% |
| 21 | 2 | **0%** | 89% | 11% | **2%** | 74% | 25% | **0%** | 80% | 20% | **0%** | 80% | 20% |
| 22 | 1 | **98%** | 2% | 0% | **92%** | 2% | 6% | **100%** | 0% | 0% | **100%** | 0% | 0% |
| 23 | 2 | **19%** | 64% | 11% | **21%** | 57% | 15% | **60%** | 20% | 20% | **100%** | 0% | 0% |
| 24 | 1 | **92%** | 6% | 2% | **98%** | 2% | 0% | **100%** | 0% | 0% | **100%** | 0% | 0% |
| 25 | 0 | **100%** | 0% | 0% | **100%** | 0% | 0% | **100%** | 0% | 0% | **100%** | 0% | 0% |
| 26 | 0 | **100%** | 0% | 0% | **100%** | 0% | 0% | **100%** | 0% | 0% | **100%** | 0% | 0% |
| 27 | 0 | **98%** | 2% | 0% | **96%** | 2% | 2% | **100%** | 0% | 0% | **100%** | 0% | 0% |
| 28 | 1 | **100%** | 0% | 0% | **100%** | 0% | 0% | **100%** | 0% | 0% | **100%** | 0% | 0% |
| 29 | 0 | **87%** | 13% | 0% | **100%** | 0% | 0% | **100%** | 0% | 0% | **100%** | 0% | 0% |
| 30 | 0 | **100%** | 0% | 0% | **98%** | 2% | 0% | **100%** | 0% | 0% | **100%** | 0% | 0% |
| Overall | | **75%** | 21% | 2% | **76%** | 20% | 3% | **77%** | 19% | 6% | **84%** | 14% | 2% |
| Text-only (0) | | **98%** | 2% | 0% | **99%** | 0% | 0% | **100%** | 0% | 0% | **100%** | 0% | 0% |
| Image-optional (1) | | **98%** | 2% | 0% | **97%** | 2% | 1% | **100%** | 0% | 0% | **100%** | 0% | 0% |
| Image-required (2) | | **36%** | 55% | 6% | **36%** | 53% | 8% | **38%** | 51% | 16% | **56%** | 38% | 5% |

*Table 2 - Performance of FCI assessment for different questions. The question types are labeled as 0 for questions with text-only, 1 for questions that include an image but can be solved without it (image-optional), and 2 for questions where the image is essential to solve the problem (image-required).*

Table 2 also summarizes the overall performance of the GPT-4o model on FCI tasks across different settings: translated versions in 53 languages, their English back-translations, and repeated runs in the original English version using GPT-4o and GPT-o1. The average performance on the translated versions using GPT-4o was 75%, and the back-translations scored slightly higher at 76%. When tested in English, GPT-4o achieved an average score of 77% across five repetitions. Additionally, the English FCIs solved with GPT-o1 outperformed all other variants with 84%. The results are further broken down by question type: text-only, image-optional, or image-required. While text-only and image-optional questions were solved with high accuracy (97–100%), performance on image-required questions was considerably lower. For example, GPT-4o's accuracy on image-required questions hovered around 36–38%, whereas GPT-o1 showed improved visual reasoning at 56%.

These results reveal a significant discrepancy between the model's proficiency in solving text versus image problems. Although GPT models demonstrate near-perfect accuracy on items relying on text, their performance on visual questions with crucial image data is more variable and substantially weaker. Therefore, to evaluate the quality of translations, our analysis focused on the text-only and image-optional questions.

Table 3 shows the results of language-specific performance grouped by the percentage of the correct answers. The "Language" column indicates the correct percentage of the translated test in the given language, and the "English back translation from" corresponds to the correct percentage of the test translated back into English.

| Correct out of 30 | Language | English back translation from |
|---|---|---|
| 19, 63% | Punjabi (3), Telugu (3) | |
| 20, 67% | Somali (1) | Somali (3) |
| 21, 70% | Urdu (2), Vietnamese (1) | Hindi, Lithuanian (1), Malayalam (2), Persian (1) |
| 22, 73% | Afrikaans, Albanian, Chinese, Czech (2), Dutch, Gujarati, Hebrew, Hindi (1), Latvian, Malayalam (1), Moldavian, Moldovan, Persian, Romanian, Russian, Slovenian (1), Spanish, Swedish, Turkish (1), Welsh | Arabic (1), Bengali (1), Czech, Finnish, Gujarati, Croatian, Italian, Kannada (2), Korean, Marathi, Norwegian, Punjabi (1), Swedish, Tamil (2), Telugu, Ukrainian, Urdu (1), Vietnamese |
| 23, 77% | Arabic, Catalan, German, Greek, Estonian, French, Indonesian, Italian, Japanese, Kannada, Korean (1), Lithuanian, Marathi, Nepali, Norwegian, Polish, Slovak, Tamil (1), Thai, Ukrainian | Albanian, Catalan, Chinese, Bulgarian, Danish, German, Greek, Estonian, French, Japanese, Latvian, Polish, Portuguese, Moldavian, Moldovan, Romanian, Russian, Slovak, Slovenian, Swahili (1), Tagalog, Turkish, Welsh |
| 24, 80% | Bulgarian, Danish, Finnish, Croatian, Hungarian, Swahili, Tagalog | Afrikaans, Dutch, Hungarian, Indonesian, Macedonian, Nepali, Thai (1) |
| 25, 83% | Bengali, Macedonian, Portuguese | Hebrew |
| 26, 87% | | Spanish |

*Table 3* - Overall performance by language. If there were any incorrect text-only or image-optional responses, the number of incorrect responses is indicated in parentheses.

As can be seen, the majority of languages perform with 70% or more correct answers and are comparable with the original English version's performance. Since this work focused on the text-only and image-optional questions and performed and in-depth analysis of the language performance of those questions, Table 3 additionally indicates the number of incorrect text-only and image-optional questions in parentheses.

Table 4 shows in detail the questions and languages of all text-only and image-optional questions that were solved incorrectly. In the discussion section, we present the particular aspects that could have affected the correctness of solutions.

| Question | Language | English back-translation from |
|---|---|---|
| 9 | Telugu, Punjabi, Tamil, Urdu | Bengali, Kannada, Malayalam, Somali, Swahili, Punjabi, Tamil, Urdu |
| 11 | Telugu | Malayalam |
| 22 | Somali | Persian, Kannada, Lithuanian, Somali |
| 24 | Czech, Korean, Malayalam, Vietnamese | Arabic |
| 27 | Punjabi | Somali, Thai |
| 29 | Czech, Hindi, Punjabi, Slovenian, Telugu, Turkish, Urdu | |
| 30 | | Tamil |

*Table 4 - Incorrect text-only or image-optional questions.*

To better understand how translation affected terminology usage across languages, we analyzed the frequency and variation of a few physics terms in both the original and back-translated versions of the FCI and identified several interesting examples. For instance, across all original English 30 FCI questions, the word "speed(s)" occurs 39 times. To make each FCI question solvable on its own, this was increased to 47. In 44 cases, "speed(s)" was used as a noun. Additionally, there were two instances of "speeds up" and one case of "speeding up". Multiplying by 53 for all translations, this leads to 2491 occurrences of the term "speed(s)", 2332 as a noun, 53 cases of "speeding up", and 106 cases of "speeds up". After translation and back-translation, the English translation contained the term "speed" 1,514 times. Amongst those cases were 10 cases of "speeds up" and two cases of "speed up". The term "velocity" was used 702 times. Additionally, all 20 instances of "kick" were translated as "push" or "hit" in Catalan, and terms like "diagram," "figure," and "image" were used interchangeably across languages. These patterns highlight how translation can subtly shift terminology, even when the intended meaning remains largely intact.

## 4. Discussion

**Transcription and Translation**

Automating the conversion of the PDF into text and PNG files using an LLM, as compared to manually cutting and pasting individual questions, saves some time but still requires multiple steps. For this study, using cut and paste would have been almost as efficient. However, we believe automated transcriptions for larger data sets can save time. The complexity of the document will determine how much manual adjustment and fixing is required. Other, more advanced models may have been more efficient and required less adjustment, but since our data did not require it, we did not explore this further.

As shown in Table 1, for some languages and questions, non-Latin letters were used by LLM in the translation of the multiple-choice selectors. Which alphabet it chose was inconsistent even within one language. For example, it used the local script commonly for Arabic or Greek (20 and 22 times, respectively). This number would have been higher if questions 6, 7, 8, 12, 14, 21, and 23 did not have Latin selectors on the picture, making translation impossible. For other languages, it happened for just a few questions, e.g., Telugu with just one question in non-Latin script.

During the problem-solving process, the individual multiple-choice options *were also* inconsistently used for questions with a non-Latin alphabet. The LLM would consistently use the language-specific alphabet for Arabic, Bulgarian, Hebrew, and Macedonian in its response. For other languages, it would answer some questions with Latin multiple-choice selectors, while others were taken from the language-specific alphabet. For example, solving the assessment in Greek, 26 multiple-choice answers were in a language-specific alphabet, which was obvious for the letters Γ or Δ, but less obvious for A, B, and E that look similar in the Greek and Latin alphabet. The LLM response was impacted by the alphabet used in the question and the instruction in the prompt. One of the prompts specifically stated to return just the correct multiple-choice letter, giving "*B*" as an example. As a result, for seven languages (Arabic, Bulgarian, Greek, Persian, Hebrew, Macedonian, and Marathi), the instruction to "restate the correct multiple-choice answer" included a local script letter instead of the Latin "*B*" in the example ("e.g. "*B*"). This increased the non-Latin responses for those languages, e.g. with Bulgarian going from two questions in non-Latin multiple-choice selector script to all responses in the Bulgarian Cyrillic alphabet. For other languages with non-Latin multiple-choice options in the questions, the non-Latin letter responses decreased if the Latin "*B*" was used in the prompt (for example, from 9 questions in Thai script to just two answers in the Thai script).

The randomness of which alphabet was chosen by the LLM during problem-solving goes even further. For example, both Marathi and Hindi use the Devanagari script and Latin alphabet for some of the multiple-choice selections. However, even when using Devanagari script, there was some randomness in how it translated from the Latin script used in English to the Devanagari script. The letter क was used at times in the correct order of the Devanagari script. It is the first letter in the alphabet and was substituted for "*A*". In other cases, it substituted for "*C*" as the Devanagari letter क is pronounced like the "ka", most similar to the letter "*C*" as in "cut" in English. Adding to the complexity of the task, a common practice in certain languages mixes scripts. For instance, Russian exams often use Latin letters while otherwise using Cyrillic script. However, the LLM was consistent with standard practice in this case, using Latin letters for the multiple-choice options while otherwise using Cyrillic script. The inconsistent mix of translations according to the order of the alphabet and phonetic conversion can cause issues when one autogrades based on the multiple-choice letter only.

Although Table 1 lists which questions had non-Latin scripts in question translations and answers, it does not show a clear correlation. This suggests that the LLM's choice of script could be influenced by the prompt formatting, the question translation itself, and the stochastic nature of the LLMs. This unpredictability can pose challenges for automated grading systems that rely solely on the multiple-choice letter.

Some physics-specific and non-physics-specific terms either lack direct equivalents in other languages or have multiple near-synonyms with subtle differences in meaning. This can lead to inconsistencies in translation and interpretation. For example, while English distinguishes between "speed" and "velocity," many languages do not maintain this distinction, which may blur important conceptual boundaries. However, for the questions in the FCI, this distinction did not matter and while notable, did not have a measurable impact on the accuracy of responses. However, there are additional examples where translation issues led to changes in the physical meaning of specific problems and those will be discussed in detail in the following section.

**Problem-solving**

The performance of the LLM varied across text-only, image-optional, and image-required question categories. Of note here is that the overall performance is significantly higher than a multilingual study of expert translations of the FCI (Kortemeyer et al., 2025). This is not unexpected as the different methodology of that study, critically, that content was provided as screenshots in a "as seen by students" manner, was more challenging for the LLM.

**Image-required questions**

For text-only and image-optional questions, the model performed exceptionally well, with some notable exceptions, which will be discussed in detail below. However, its performance dropped to 36% for image-required questions. Detailed review of the image-required questions shows recurring issues: misinterpreting arrows (Q7), confusing curved and straight trajectories (Q8, Q12, Q14), and difficulty counting discrete points (Q19, Q20). Even when the model demonstrates sound conceptual reasoning, it frequently fails to select the correct visual representation, highlighting its limited spatial reasoning abilities, a known limitation of GPT-4 Vision that aligned with previous research (Polverini & Gregorcic, 2024a). Despite this overall trend, three image-required questions, Q5, Q6, and Q18 had a high correct answer rate, suggesting that specific visual features may be more accessible to the model than others.

From an educational standpoint, these weaknesses may be beneficial by discouraging overreliance on AI for visual problem-solving. Since the model cannot reliably interpret complex images, students are less likely to bypass genuine understanding in favor of automated answers. This opens avenues for integrating AI into physics education as a conceptual aid while maintaining assessment integrity. Enhanced visual reasoning in AI models is necessary to improve their value in diverse educational contexts, particularly where language barriers hinder comprehension.

**Text-only or image-optional questions**

There were 19 questions that were either text-only (10 questions) or image-optional (9 questions). Before translation, both GPT-4o and GPT-o1 answered all those questions correctly. With five repetitions each, that is 190 correct answers without a mistake. For these questions that the LLM consistently answers correctly in the original English version, it may answer them incorrectly after translation due to a few reasons:

1. The statistical nature of LLMs,
2. Issues with the translation into a non-English language,
3. Issues with solving the question in a non-English language,
4. Issues with the translation back to English.

In some cases, it may be a combination of these reasons. For example, a translation may be imprecise, causing it to have a higher statistical likelihood of getting a question incorrect. Due to the required computing time and cost, there was only one translation done into each language. More rigorous statistical analysis could be done in isolating individual languages with more samples in a future study.

For the 53 languages, there are 1007 responses after translation to questions that did not require an image. Of those, 18 were counted as incorrect. For the 1007 responses back-translated to English, 17 questions were counted as incorrect. Dividing the assessment item into two parts: question and multiple-choice answer options, we can separate two main translation mistake categories. Changes in the question itself may result in the fact that a different option becomes correct or none of the options are correct. Certain changes in the multiple-choice options result in contradictory or subtle implications leading to wrong answers. Looking at these questions in more detail provides an insightful illustration of what issues can occur when translating from one language to another.

**Translation Issues with Technical Physics Terms**

Accurate translation of technical terminology is critical for ensuring the integrity of multilingual assessments in physics. Our analysis revealed several instances where translation errors led to discrepancies in meaning, ultimately affecting the LLM responses. Looking at the back-translations for question 9, the original phrase "arithmetic sum" was incorrectly translated as "vector sum" in two of the multiple-choice options in the following languages: Bengali, Kannada, Malayalam, Punjabi, Swahili, Tamil, Telugu, and Urdu. This completely changed the meaning of the question and a different multiple-choice option was correct. The Bengali, Kannada, Malayalam, Somali, and Swahili translations still had the correct response and it appears the issue happened with the back-translation to English. From Table 4, for Punjabi, Tamil, and Urdu, we conclude that the error already occurred in the translation from English. For Telugu, we conclude that the LLM translated incorrectly to Telugu and then back-translated incorrectly to English, fixing the first incorrect translation. Despite these translation issues, the LLM selected the correct answer based on the translated choices in all languages. This indicates that the model understood the question correctly and that the error lies in the translation process.

For question 22 in Somali, the translation incorrectly states "perpendicular" to "opposing" in the question prompt. The LLM then answered the question based on a force acting in a different direction and picked the correct answer for this arrangement but incorrect multiple-choice option as originally written.

For question 29, in Hindi, the LLM used a nonsensical translation of "net-force" that would translate as something like "pure-force" into English. The LLM translation back to English had the correct term "net force" and only with the help of a Hindi-speaking physicist was this issue identified. Interestingly, for Kannada and Malayalam, translation back to English used exactly this term "pure force", though in these cases, the LLM answers were correct. Clearly, unless a physicist with appropriate language skills checks the translations, one cannot be sure of the full accuracy of an LLM translated physics text, even if the LLM answers a question correctly and the control translation back to English is accurate.

**Critical Meaning Changes in Physics Context**

In several cases, translated versions introduced terms or phrasing that altered the physical assumptions of the scenario. For question 27, in Somali, the translation back to English had some changes, for example, "*speed*" became "*velocity*", the correct choice "immediately start slowing to a stop" became "immediately start to slow down", which would not have altered the meaning of the question for a physicist. However, it added one crucial element. The "*horizontal floor*" in the initial question became "*smooth horizontal floor*". For a physicist, this would lead to a contradiction in the question. On one hand, the question required friction as the speed stayed constant even though an external force was applied. On the other hand, the word "*smooth*" implied that there was no friction. We speculate that the LLM has seen so many physics problems with smooth frictionless floors that it hallucinated the smooth into the question. Once that was there, the response had to deal with contradictory information. A previous study has shown that reconciling those contradictions in physics is something GPT-4 struggles with (Dunlap et al., 2025). In Thai, the correct choice "immediately start slowing to a stop" became an incorrect option "Start slowing down and stop immediately". This left the LLM with no correct answer to choose from. It chose the arguably closest incorrect answer that indicated the object slowed down, though this option still included an incorrect statement that the box would continue to move at a constant speed for a while. Like before, the LLM was not able to point out contradictions in the question. In addition to the answers in English back-translations from Somali and Thai, the answer for Punjabi was incorrect. Looking at the translated response, it seems it had a similar issue to the Thai above. In the English translation of the response, it specifically states correctly, albeit slightly awkwardly, that "*The box will not stop immediately because friction gradually reduces motion.*" It seems the correct choice was altered again, leaving the LLM no correct choice. While we did not have a Punjabi-speaking physicist in our group, this is supported by the translation of the question back to English as "*Will gradually slow down to a stop immediately*." Interestingly, it chose the correct option in the English translation. Here, it made contradictory statements in the answer, stating at one point that "*The box will not stop immediately, as friction causes a gradual deceleration.*", but then selected the option that had it stop immediately without strongly pointing out the contradiction.

For question 30, in the English back-translation from Tamil, the LLM response used the correct reasoning and selected an accurate response to a slightly, yet from a physics perspective, significantly altered question. The translation had changed "*Despite a very strong wind*" to "*Ignoring the strong wind*" and GPT-4 then selected the choice that indeed ignored the force exerted by the wind.

**Subtle Interpretive and Contextual Issues**

Subtle changes in the back-translation of question 11 from Malayalam to English resulted in the LLM ignoring the offsetting forces of gravity and the normal force from the surface. It concluded correctly that there are no forces in the horizontal direction, but because the translation incorrectly implied that only forces acting in the direction of the motion mattered, it ignored the vertical forces. The answer in Telugu correctly lists the gravitational and normal force, but incorrectly argues that there remains a horizontal force in the direction of the swift kick. It refers in its response to a horizontal arrow in the accompanying image and it is possible that the image interferes with the text-based instruction.

The response to question 24 for the back-translation from Arabic relied heavily on the image. It appears that even though one could have answered the question without the image, it heavily influenced the LLM's response and it led it astray. In Korean, it provides two answers. The correct one with the qualifier that this assumes that the thrust has stopped for the problem in question. The incorrect answer states that it is valid, assuming that it continued. In Malayalam and Czech, it assumes that the thrust remained constant. Translating the Vietnamese answer shows that the physics reasoning is accurate for this language. However, it picked the multiple-choice option that would include the motion prior to the point asked in the question. If given by a student, most instructors would have probably given full credit if the answer was a long form open response problem, but the multiple-choice selection was incorrect, nevertheless.

For question 29, the incorrect multiple-choice answer was given for Czech, Hindi, Punjabi, Slovenian, Telugu, Turkish, and Urdu, but was then correct in all the back-translations to English. The translated questions and responses were reviewed by a native speaker for Czech and Hindi and additionally, back-translations were reviewed in English for all seven incorrect responses. In the original English version of this question, the force exerted by air is not relevant. However, the translated versions of the questions seem to subtly imply in the multiple-choice option that the force due to air is not just a force that was provided as an incorrect option but instead has to be considered since it was included in the question. For example, in Czech, the LLM response reads "*Resultant air force (downward): The effect of air on the chair is very small and usually negligible. However, here it is explicitly mentioned that this force acts. Therefore, we must include it in the answer. Even though it is small, it is still real and acts downward.*" A Czech translation of the FCI done by a physicist is available on PhysPort. When given the physicist's translation of the question, GPT-4o answered correctly 3 out of 3 times. Overall, the performance of the Czech FCI translated by LLM is 73% with only one incorrect text-only response. However, the LLM's correct responses to the physicist's translation of question 29 and the incorrect interpretation of the LLM-translated question suggest there are still cases where subtle phrasing and word choices done by an expert are superior.

5. **Limitations, implications, and further questions**

The findings of this study offer promising insights but also raise important questions about the robustness, reliability, and broader applicability of LLMs in physics education. Our results must be interpreted in light of specific limitations, technical challenges, and open questions that remain unresolved. In the following, we reflect on these aspects, ranging from system parameters and translation precision to the interpretation of images and the broader implications for educational equity and safety.

First, all translations and responses were generated using the default temperature setting of the LLM. While this choice reflects typical user behavior and ensures consistency, it is possible that other settings could have produced different results. Additionally, only one translation was generated for each of the 53 languages due to computational constraints, limiting the statistical robustness of our conclusions. The detection of translation errors was further complicated by the lack of fluency in all tested languages among the research team. Some inaccuracies may have gone unnoticed, particularly in cases where the back-translated English appeared correct despite issues in the non-English version. Moreover, certain terms in physics, similar to the earlier discussed "velocity" versus "speed", may collapse into a single word in other languages (e.g.,

Geschwindigkeit in German), leading to subtle but consequential shifts in meaning. The model's limited ability to interpret visual information remains a bottleneck, despite recent progress, and raises concerns about its suitability for image-based problem-solving (Polverini & Gregorcic, 2024a). Finally, the possibility that the FCI assessment questions and answer key were present in the model's training data cannot be fully excluded; however, the observed performance shifts in response to translation errors suggest that model outputs are still sensitive to semantic changes, not just memorized content. More advanced multimodal reasoning models like GPT-o3, GPT-o4-mini, or GPT-5 may further improve image interpretation capabilities. However, clearly, further improvements are necessary for AI to deal with the physics interpretation of image data.

Our study focuses on transferability across languages and text-based physics reasoning. With more than 98% correct responses, the LLM's ability to both translate the FCI into a large number of languages without changing how it interprets its physics content is remarkable. While impressive, this high but not perfect accuracy comes with some important implications. Our study indicates that we are already at a point where language-agnostic physics delivery of physics content seems viable. However, LLMs still make rare translation errors, and some of them can go beyond style and completely change the meaning of a physics problem. Given the stochastic nature of LLMs, eliminating the rare mistakes may prove to be quite challenging. One might argue that these rare errors are acceptable for low-stakes educational materials. However, even a low-stakes homework assignment that was incorrectly translated could be frustrating for students. It gets even more problematic for higher-stakes tasks like exams. The stakes are raised further for real-world applications, for example, in engineering. Some 25 years ago, NASA lost a Mars orbiter using English units instead of metric units (NASA Safety Center, 2009). If we deal with language instead of scientific units, how do we prevent a multilingual team using AI to translate materials from making similar errors? This time, based on a rare but important translation error. At one point, a physics expert with language knowledge may still be required. Training students with the skills to identify rare errors by AIs is a new challenge for educators going forward. It seems both traditional content knowledge and new skills to use AI efficiently and responsibly are important here. This is particularly challenging considering that LLMs can perform better than almost all students on text-based introductory-level physics tasks.

## 6. Conclusion

The study on translating the FCI using OpenAI's GPT-4o highlights the transformative potential and challenges of AI-driven multilingual educational tools. Understanding how LLM handles language-specific physics assessments will provide insights into its applicability for global education. One of the main outcomes is the realization that while LLMs can greatly facilitate the translation process, enhancing access to physics education across different languages, certain subtleties and context-specific terms in scientific language pose significant challenges. This can lead to mistranslations or subtle nuances being lost. Despite these challenges, the LLM demonstrated strong problem-solving capabilities of translated and back-translated inventory questions, providing insight into the underlying proficiency and limitations of AI in educational contexts.

Emphasizing the importance of precision, the study finds that translation errors, particularly those involving technical terminology, could alter the intended meaning of physics problems. These errors underline the necessity for human oversight, ensuring the accuracy and pedagogical integrity of translated content. Although LLMs performed efficiently in text-based questions and displayed sophisticated reasoning capabilities, discrepancies in interpreting visual elements highlight a common limitation of LLMs, drawing attention to the need for continuous improvement in multimodal capabilities of AI systems.

Ultimately, the study illustrates that while AI could offer substantial benefits in democratizing access to educational materials, attention to quality assurance processes remains essential. By analyzing LLM performance across multiple languages, this study contributes to the growing discourse on AI's role in education, particularly in STEM disciplines. Given the increasing reliance on AI for educational purposes, it is essential to assess its limitations and strengths in multilingual environments. The results of this study provide practical insights for physics educators, policymakers, and AI developers seeking to improve the accessibility and reliability of educational resources worldwide. We give pointers on how physics educators can use AI to create their own translations of physics content and discuss challenges and opportunities going forward.

bias in multilingual language models: Cross-Lingual transfer of debiasing techniques*. arXiv.Org. https://arxiv.org/abs/2310.10310

Shahriar, S., & Hayawi, K. (2023). Let's have a chat! A conversation with chatgpt: Technology, applications, and limitations. *Artificial Intelligence and Applications*, *2*(1), 11–20. https://doi.org/10.47852/bonviewaia3202939

Singh, G., Gupta, A., Verma, P., Chaudhary, N., Kler, R., & Thakur, A. (2024). Catalyzing multilingual NLP: New methods for low-resource language support. *2024 4th International Conference on Technological Advancements in Computational Sciences (ICTACS)*, 67–75. https://doi.org/10.1109/ictacs62700.2024.10840816

Steigerwald, E., Ramírez-Castañeda, V., Brandt, D. Y. C., Báldi, A., Shapiro, J. T., Bowker, L., & Tarvin, R. D. (2022). Overcoming language barriers in academia: Machine translation tools and a vision for a multilingual future. *BioScience*, *72*(10), 988–998. https://doi.org/10.1093/biosci/biac062

Steinert, S., Avila, K. E., Ruzika, S., Kuhn, J., & Küchemann, S. (2024). Harnessing large language models to develop research-based learning assistants for formative feedback. *Smart Learning Environments*, *11*(1). https://doi.org/10.1186/s40561-024-00354-1

Steinert, S., Krupp, L., Avila, K. E., Janssen, A. S., Ruf, V., Dzsotjan, D., Schryver, C. D., Karolus, J., Ruzika, S., Joisten, K., Lukowicz, P., Kuhn, J., Wehn, N., & Küchemann, S. (2024). Lessons learned from designing an open-source automated feedback system for STEM education. *Education and Information Technologies*. https://doi.org/10.1007/s10639-024-13025-y

Stoen, S. M., McDaniel, M. A., Frey, R. F., Hynes, K. M., & Cahill, M. J. (2020). Force Concept Inventory: More than just conceptual understanding. *Physical Review Physics Education Research*, *16*(1). https://doi.org/10.1103/physrevphyseducres.16.010105

Tao, Y., Viberg, O., Baker, R. S., & Kizilcec, R. F. (2024). Cultural bias and cultural alignment of large language models. *PNAS Nexus*, *3*(9). https://doi.org/10.1093/pnasnexus/pgae346

Weglot & Nimdzi Insights. (2022). *The State of Machine Translation for Websites. A Comparative Study of the Top Machine Translation Engines*.

West, C. G. (2023, March 2). *AI and the FCI: Can chatgpt project an understanding of introductory physics?* arXiv.Org. https://arxiv.org/abs/2303.01067

Yeadon, W., & Hardy, T. (2024). The impact of AI in physics education: A comprehensive review from GCSE to university levels. *Physics Education*, *59*(2), 025010. https://doi.org/10.1088/1361-6552/ad1fa2

Yusuf, A., Pervin, N., & Román-González, M. (2024). Generative AI and the future of higher education: A threat to academic integrity or reformation? Evidence from multicultural perspectives. *International Journal of Educational Technology in Higher Education*, *21*(1). https://doi.org/10.1186/s41239-024-00453-6
**Statements and Declarations**

**Acknowledgments**

The authors thank Priya Jamkhedkar for her help with the Marathi and Hindi translations.22

**Competing Interests**

The authors have no relevant financial or non-financial interests to disclose.

**Author Contributions**

All authors contributed to the study conception and design. The research idea, code execution, and numerical data acquisition were performed by Ralf Widenhorn. Data analysis, literature review, manuscript drafting, data verification, and interpretation were carried out by Marina Babayeva. Justin Dunlap contributed to manuscript development, data interpretation, quality control, and proofreading. Marie Snětinová contributed to data analysis, data interpretation, and proofreading. All authors commented on previous versions of the manuscript and approved the final version.

**Data availability**

The data supporting the findings of this study are available upon request from the authors.